# E-MIM: Enhanced Masked Image Modeling to Avoid Model Collapse on Multi-modal MRI Datasets

Linxuan Han, Sa Xiao, Zimeng Li, Haidong Li, Xiuchao Zhao, Yeqing Han, Fumin Guo, Xin Zhou

*Abstract*—Multi-modal magnetic resonance imaging (MRI) provides information of lesions for computer-aided diagnosis from different views. Deep learning algorithms are suitable for identifying specific anatomical structures, segmenting lesions, and classifying diseases with magnetic resonance images. Manual labels are limited due to the high expense, which hinders further improvement of accuracy. Self-supervised learning, particularly masked image modeling (MIM), has shown promise in utilizing unlabeled data. However, we find model collapse problem when applying MIM to multi-modal MRI datasets. The performance of downstream tasks does not see any improvement following the collapsed model. To solve model collapse problem, we analyze and address it in two types: complete collapse and dimensional collapse. We find that complete collapse occurs because the collapsed loss value in multi-modal MRI datasets falls below the normally converged loss value. Based on this, the hybrid mask pattern (HMP) masking strategy is introduced to elevate the collapsed loss above the normally converged loss value and avoid complete collapse. Additionally, we reveal that dimensional collapse stems from insufficient feature uniformity in MIM. We mitigate dimensional collapse by introducing the pyramid barlow twins (PBT) module as an explicit regularization method. Overall, we construct the enhanced MIM (E-MIM) with HMP masking strategy and PBT module to avoid model collapse during pretraining on multi-modal MRI. Experiments are conducted on three multi-modal MRI datasets (BraTS2023, PI-CAI, and LungasMRI) to validate the effectiveness of our approach in preventing both types of model collapse. By preventing model collapse, the training of the model becomes more stable, resulting in a decent improvement in performance for downstream segmentation and classification tasks. The code is available at https://github.com/LinxuanHan/E-MIM.

This work was supported by National key Research and Development Program of China (2022YFC2410000), National Natural Science Foundation of China (82127802, 21921004), Key Research Program of Frontier Sciences (ZDBS-Y-JSC004), the Strategic Priority Research Program of the Chinese Academy of Sciences (XDB0540000), Hubei Provincial Key Technology Foundation of China (2021ACA013, 2023BAA021), and Natural Science Foundation of Hubei Province (2023AFB1061).

L. Han is with the Key Laboratory of Magnetic Resonance in Biological Systems, State Key Laboratory of Magnetic Resonance and Atomic and Molecular Physics, National Center for Magnetic Resonance in Wuhan, Wuhan Institute of Physics and Mathematics, Innovation Academy for Precision Measurement Science and Technology, Chinese Academy of Sciences-Wuhan National Laboratory for Optoelectronics, Huazhong University of Science and Technology, Wuhan 430071, China (e-mail: hanlinxuan@hust.edu.cn).

S. Xiao, Z. Li, H. Li, X. Zhao, and Y. Han are with the Key Laboratory of Magnetic Resonance in Biological Systems, State Key Laboratory of Magnetic Resonance and Atomic and Molecular Physics, National Center for Magnetic Resonance in Wuhan, Wuhan Institute of Physics and Mathematics, Innovation Academy for Precision Measurement Science and Technology, Chinese Academy of Sciences, Wuhan 430071, China, also with the University of Chinese Academy of Sciences, Beijing, 100049, China (e-mail: xiaosa@wipm.ac.cn; zm_li@hust.edu.cn; haidong.li@wipm.ac.cn; xiuchao.zhao@apm.ac.cn; hanyeqing@wipm.ac.cn).

F. Guo is with Wuhan National Laboratory for Optoelectronics, Department of Biomedical Engineering, Huazhong University of Science and Technology, Wuhan 430074, China (e-mail: fguo@hust.edu.cn).

X. Zhou is with the Key Laboratory of Magnetic Resonance in Biological Systems, State Key Laboratory of Magnetic Resonance and Atomic and Molecular Physics, National Center for Magnetic Resonance in Wuhan, Wuhan Institute of Physics and Mathematics, Innovation Academy for Precision Measurement Science and Technology, Chinese Academy of Sciences, Wuhan 430071, China, also with the University of Chinese Academy of Sciences, Beijing, 100049, China, and also with Key Laboratory of Biomedical Engineering of Hainan Province, School of Biomedical Engineering, Hainan University, Haikou 570228, China (e-mail: xinzhou@wipm.ac.cn). (Corresponding author: Xin Zhou.) (Linxuan Han and Sa Xiao contributed equally to this work.)

## I. INTRODUCTION

MULTI-MODAL medical image analysis is essential in medical diagnosis due to the progress in multimedia technology. Among them, multi-modal magnetic resonance imaging (MRI) can provide complementary and mutually informative data about tissue composition (i.e., anatomical or functional information), enabling intuitive insight into the human body's interior. This assists radiologists and clinicians in detecting and treating diseases more efficiently. For example [1], to accurately define the shapes, sizes, and diffused locations of gliomas, the whole tumor (WT) region is highly distinguishable with fluid attenuation inversion recovery (FLAIR), and the enhancing tumor (ET) region exhibits clear structures in T1-weighted contrast-enhanced imaging (T1ce). In recent years, numerous deep learning-based computer-aided diagnosis algorithms have been proposed to fuse modalities and extract features for MRI segmentation and classification. Additionally, these data-driven algorithms, such as U-Net [2] and its variants [3]-[8], require a large dataset with accurate annotations. Medical image annotations are still manually performed by neuro-radiologists, which is extremely tedious and time-consuming. Moreover, acquiring annotations for multi-modal MRI poses greater difficulty compared to single-modal MRI. This underscores the urgent need to resolve the issue of effectively leveraging unlabeled data for model training.

Self-supervised learning (SSL) is a viable paradigm for learning effective visual features without manual annotations on multimedia analysis [10]-[13]. In other words, SSL shows potential for boosting the performance of multi-modal MRI segmentation and classification by using unlabeled data. During the self-supervised training phase, a pretext task is designed for the model to solve. When trained with pretext tasks, the model can learn kernels that capture low-level and high-level features. After the self-supervised training is completed, the learned features can be further transferred to downstream tasks (such as segmentation and classification) as pretrained models, improving the performance of downstream tasks.

Based on different pretext tasks, SSL can be categorized into two main types: contrastive learning (CL) [11]-[15] and masked image modeling (MIM) [16]-[24]. CL learns effective features by pulling semantically similar (positive) samples together and pushing different (negative) samples in the latent space. The pretext task of CL defines an image and its augmented images as positive samples, with all other images considered as negative samples. Effective features learning in CL heavily rely on image augmentation [11]. However, certain augmentation methods, including scaling, rotation, affine, and deformed transformations, are insufficient when applied to medical image datasets [25]. On the other hand, MIM does not implement any augmentation methods during the self-supervised training phase [19]. The pretext task of MIM focuses on reconstructing the missing patches that are randomly masked. MIM has achieved state-of-the-art performance across different datasets.

However, we find model collapse when using MIM for pretraining on multi-modal MRI datasets.

Model complete collapse is observed first. The completely collapsed model only outputs trivial solutions for any input [15]. Transferring a completely collapsed model does not lead to any improvement in performance for downstream tasks. Fig. Fig. 1 (a) illustrates the results of reconstructing masked images using MAE (a type of MIM) on the BraTS2023 dataset. For any input, the masked pixels reconstructed by MAE are the values lacking semantic information, which were proven in Section III-A to be the average values (i.e., trivial solutions) of the positions of these pixels on the entire dataset. This indicates the existence of model complete collapse. In this work, we investigate the mechanism of model complete collapse from the perspective of the loss function. The loss of a completely collapsed model is the variance of the masked images in the entire dataset. By conducting a Monte Carlo simulation, we discover that the variance of the masked images in multi-modal MRI datasets is much lower than the converged loss value in a normally trained model that is not completely collapsed. It means the model parameters will not be optimized through learning effective features but will be optimized into a completely collapsed model. Based on this, we propose a hybrid mask pattern (HMP) masking strategy to directly increase the variance of masked images in multi-modal MRI above the converged loss value of a normally trained model. This approach ensures that when the model falls into complete collapse during training, the loss has not yet converged. As a result, the parameters of the model can still be optimized along the gradient, allowing it to learn semantic features for reconstructing masked images and avoiding complete collapse.

The model still suffers from dimensional collapse when only the model complete collapse is addressed. Model dimensional collapse refers to the situation where the learned features are in low dimensional subspace [23]. Even though a dimensionally collapsed model can extract a few features, transferring it still does not lead to any improvement in performance for downstream tasks. The singular values [26] and effective rank [27] of the learned feature illustrate the existence of model dimensional collapse. As depicted in Fig. 1 (b), the effective ranks of the learned features decrease progressively during the learning process. Fig. 1 (c) demonstrates that after learning, features learned by MAE become more collapsed compared to those from random initialization, as there are fewer large singular values. In this work, we provide a theoretical explanation of the MIM mechanism from the perspective of the joint probability distribution of image patches. Essentially, MIM implicitly aligns groups of similar image patches through randomly masking. However, it falls short in ensuring feature uniformity, resulting in a few effective features learned and leading to model dimensional collapse. To address this issue, we introduce the pyramid barlow twins (PBT) module as an explicit regularization method to improve feature uniformity, increase the number of effective features learned, and prevent model dimensional collapse.

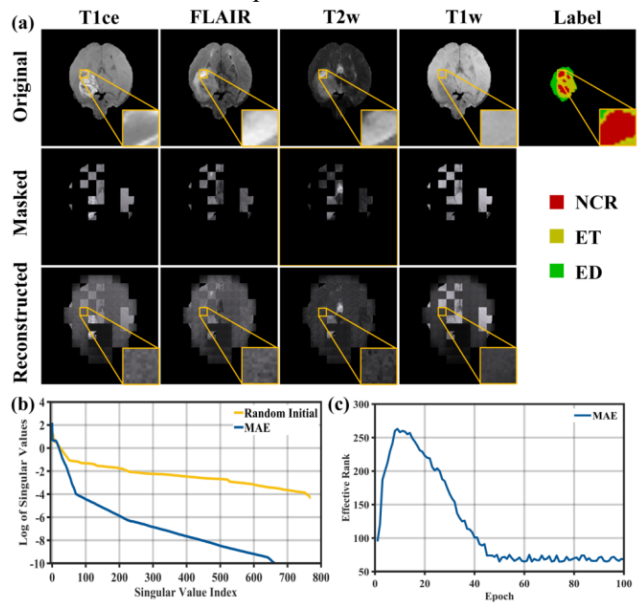

**Fig. 1.** (a) Example result of one MRI scan from the BraTS2023 validation set. As the raw images are all 3D volumes, images are shown in the form of slices. The necrotic tumor core (NCR), enhancing tumor (ET), and edematous tissue (ED) are presented in red, yellow, and green, respectively, in the last column. Each row corresponds to one example. For each modal, the original images (top), the masked images (middle), and the MIM reconstruction patches (bottom) are shown. (b) Comparison of the singular values of learned features with MAE and randomly initialized parameters. (c) The changing process of the effective rank of the encoded features trained with MAE.

Based on the information provided, the proposed Enhanced MIM (E-MIM) merges the HMP masking strategy and the PBT module to avert model collapse during pretraining on multi-modal MRI dataset. The study is performed on three multi-modal MRI datasets: BraTS2023, PI-CAI, and LungasMRI. The results demonstrate the efficacy of our approach in avoiding both complete and dimensional collapse. The transfer of the pretrained model to segmentation and classification tasks lead to substantial enhancements in result accuracy.

In summary, our contributions are as follows:

1) To the best of our knowledge, this is the first time that model collapse has been achieved by pretraining MIM on multi-modal MRI datasets. We have also proposed an enhanced masked image modeling, dubbed E-MIM, to avoid model collapse from complete collapse and dimensional collapse.

2) We proposed hybrid mask pattern to directly increase the collapsed loss value and surpass the normally converged loss value. This enables the model parameters to be continuously optimized for a normal (non-collapsed) result, thereby avoiding complete collapse.

3) We introduce the pyramid barlow twins module, which applies the cross-correlation matrix-based barlow twins loss on different vision scales. This explicitly enhances feature uniformity to prevent dimension collapse.

4) Experiments on three multi-modal MRI datasets demonstrate that the proposed method can prevent complete collapse and dimensional collapse. Pretrained models using the proposed method are transferred to segmentation and classification tasks, effectively improving the accuracy of the results.

## II. RELATED WORK

### A. Multi-modal MRI Analysis

So far, many deep learning based multi-modal MRI analysis methods have been proposed. For instance, Schelb et al. [28] trained a U-Net with T2-weighted and diffusion MRI to segment prostate cancer. Based on U-Net, several networks, such as TransUNet [3], [29] and UNETR [4] employing vision transformer (ViT), have been proposed to improve the segmentation performance on BraTS2020. Yang et al. [30] designed a dual disentanglement network for brain tumor segmentation with missing modalities. Chen et al. [32] proposed a modal-specific information disentanglement framework to extract inter- and intra-modal attention maps on multi-modal MRI. Liang et al. [32] employed a 3D Dense Net model to predict Isocitrate Dehydrogenase (IDH) genotypes with all tumor lesion patches. Furthermore, Cheng et al. [33] combined the tasks of glioma genotyping and segmentation into a multi-task network on BraTs2020. In addition, Li et al. [34] proposed a complementation-reinforced network to reconstruct and segment pulmonary gas MRI.

### B. Self-supervised Learning

Supervised learning methods mentioned above rely on fully annotated medical datasets. As a viable alternative, self-supervised learning obtains supervisory signals from the given data and learns generalizable dense representations of the input. There are two directions in self-supervised learning.

The first direction is contrastive learning [9], [35]-[37] which defines positive and negative sample pairs as learning tasks and treats them differently in the loss function. Marin et al. [35] pretrained the ResNet50 backbone for object detection in chest X-ray. Zheng et al. [9] proposed a multiscale visual representation learning algorithm to perform finer-grained representation and to handle different target scales for downstream segmentation tasks. Together with the contrast loss, Zhou et al. [36] presented preservation contrastive learning for self-supervised medical representations. Moreover, Jiang et al. [37] designed a conditional anatomical feature alignment module to complement corrupted embeddings with globally matched semantics to create contrastive pairs in 3D medical analysis.

The other direction is MIM [19], [21], [24], [38], [39], which learns representations by recovering masking-corrupted images. MAE [19] demonstrates, for the first time, that masking a high proportion of the input images can yield a non-trivial and meaningful self-supervisory task. It has some advantages, such as working with minimal or no augmentation. It represents the state-of-the-art (SOTA) on natural datasets. Yan et al. [38] proposed a privacy-preserving and federated self-supervised learning framework that collaboratively train models on decentralized data using masked image modeling as the self-supervised task. Chen et al. [21] employed masked image modeling approaches to advance 3D medical image analysis using a lightweight decoder for reconstruction. Haghighi et al. [39] developed a framework that unites discriminative, restorative, and adversarial learning in a unified manner. Yan et al. [24] proposed a hybrid visual representation learning framework for self-supervised pretraining on large unlabeled medical datasets using contrastive and generative modeling.

### C. Model Collapse

Self-supervised learning, which aligns similar data points closely in the representation space, may result in model collapse [15], hindering any performance enhancement in subsequent tasks. Model collapse can be shown as complete collapse, where the model only produces trivial solutions, and dimensional collapse, where learned features are confined to a low-dimensional subspace [26].

Some studies have introduced new self-supervised frameworks to avoid model collapse. For example, LeCun et al. [20] measured the cross-correlation matrix between the outputs of two identical networks called Barlow Twins. Zhang et al. [40] presented a new approach termed Align Representations with Base, which aligns the learned embeddings to intermediate variables. Compared to Barlow Twins, ARB does not require pairwise decorrelation, resulting in linear complexity. Also, some researchers attempted to avoid model collapse by redesigning proxy tasks and loss function. Wang et al. [41] employed a network ended with a SoftMax operation to produce Twins-class distributions of two augmentation images. Moon et al. [42] used the same random vector for the augmented embeddings of the given image. This implies that the embeddings are locally dispersed, giving a latent contrast effect between the embeddings of different images. Zhang et al. [23] established a close connection between MAE and contrastive learning and showed that MAE implicitly aligns the mask-induced positive pairs. They proposed a uniformity-enhanced MAE loss that can address model collapse.

## III. METHOD

In Section III-A, a theoretical explanation is provided for the model complete collapse observed when employing MIM pretraining on multi-modal MRI. Expanding on this explanation, the hybrid mask pattern is introduced, aimed at avoiding model complete collapse by directly enhancing the lower bound of the collapsed loss value above the converged loss value of a normally trained model. Section III-B delves into the operational a mechanism of MIM from the perspective of image patches, revealing that MIM focuses on feature alignment while lacking feature uniformity. This results in the learned features being in a low-dimensional subspace, which consequently causes model dimensional collapse. To address this, a pyramid barlow twins module is proposed as a means of explicit uniformity regularization to promote feature uniformity and avoid model dimensional collapse. In Section III-C, the hybrid mask pattern is integrated with the pyramid barlow twins module, offering a detail of the structure and loss function of the enhanced MIM proposed. The hybrid masking pattern is utilized for masking the input image, while the pyramid barlow twins module serves the purpose of explicitly regularizing the encoder during pre-training. Through the effective combination of these two modules, E-MIM addresses challenges associated with complete and dimension collapse.

### A. Avoiding Complete Collapse by Hybrid Mask Pattern

Firstly, the mathematical formulation of MIM training on multi-modal MRI is introduced. Given a volume $x \in \mathbb{R}^{C \times H \times W \times D}$ $H \times W \times D$ and $C$ modalities from a multi-modal MRI dataset. The volume $x$ is divided into n patches, denotes as $x = \{p^1, p^2, \cdots, p^n\}$, $p \in \mathbb{R}^{C \times 16 \times 16 \times 16}$ [8]. Then, a binary mask $m \in \{0,1\}^n$ with a masking ratio $\rho$ (i.e. the ratio of removed patches) is randomly drawn on the patches. For clarification, $u$ and $m$ denotes unmasked and masked patches in $\{p^1, p^2, \cdots, p^n\}$, respectively. Unmasked view is $x_u = \{u^1, u^2, \ldots, u^{n*(1-\rho)}\}$ and the masked view is $x_m = \{m^1, m^2, \ldots, m^{n*\rho}\}$, whose marginal distributions are $\mathcal{M}(x_u | x) = \mathcal{M}(u^1, u^2, \ldots, u^{n*(1-\rho)} | x)$ and $\mathcal{M}(x_m | x) = \mathcal{M}(m^1, m^2, \ldots, m^{n*\rho} | x)$. The joint distribution is $\mathcal{M}(x_u, x_m | x)$, where $x_u$ and $x_m$ are complementary, i.e. $x = x_u \cup x_m$.

MIM model is an encoder-decoder architecture, denoted as $h = g \circ f$. The encoder $f$ maps inputs $x_u$ to a latent feature $z = f(x_u)$, and the decoder $g$ maps $z$ to reconstruct the $x_m$. The loss of MIM is the mean square error loss:

$$\mathcal{L}_{MIM} = \mathbb{E}_x \mathbb{E}_{x_u, x_m | x} \|g(f(x_u)) - x_m\|^2 \quad (1)$$

If the model is complete collapse, it only outputs trivial solutions for any input, i.e. $\forall x_u, g(f(x_u)) = c$. Then,

$$\mathcal{L}_{MIM} = \mathbb{E}_{x_m} \|c - x_m\|^2 \quad (2)$$

when c is the mean value of $x_m$, $c = \mathbb{E}_{x_m} x_m$, the $\mathcal{L}_{MAE}$ reaches its minimum value:

$$\mathcal{L}_{MIM} \geq \mathbb{E}_{x_m} \|\mathbb{E}_{x_m} x_m - x_m\|^2 = Var(x_m) \quad (3)$$

where $Var(x_m)$ is the variance of masked view $x_m$. In other words, when the model is complete collapse, $Var(x_m)$ is the lower bound of $\mathcal{L}_{MIM}$. Due to the diversity in the nature image datasets training on MIM, such as CIFAR-10, the $Var(x_m)$ will be high, making it difficult to minimize the MAE loss to a small value with a collapsed model.

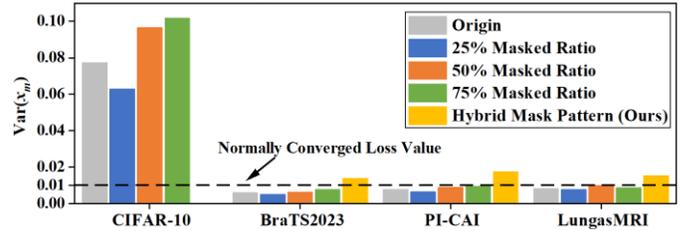

**Fig. 2.** Result of Monte Carlo sampling of calculating $Var(x_m)$ with different mask ratios and Hybrid Mask Pattern on the CIFAR-10, BraTS2023, PI-CAI, and LungasMRI datasets. The normally converged loss value is the empirical value obtained from a normally trained MIM on CIFAR-10.

Nevertheless, multi-modal MRI datasets like BraTS2023, PI-CAI and LungasMRI exhibit limited image diversity. This is primarily due to the consistent acquisition of images using fixed sequences, resulting in a similar style. Additionally, all images are registered to a common atlas, leading to a same shape for each image. While variations may exist in the lesions or organs depicted, they typically occupy a small portion of the overall image. To verify the $Var(x_m)$ of different datasets, Monte Carlo sampling is used to calculate the $Var(x_m)$ under CIFAR-10, BraTS2023, PI-CAI and LungasMRI datasets with different masking ratios $\rho$, as shown in Fig. 2. The normally converged loss value (approximately 0.01) is obtained from trained MIM on CIFAR-10, where the MIM is not collapsed. The $Var(x_m)$ on CIFAR-10 is much higher than that on the BraTS2023, PI-CAI, and LungasMRI datasets. It can also be observed from the figure that the $Var(x_m)$ in CIFAR-10 is much higher than the normally converged loss value. In multi-modal MRI datasets, $Var(x_m)$ is lower than the normally converged loss value. In other words, the collapsed loss value of multi-modal datasets is lower than the normally converged loss value. This causes the parameters to not be optimized to learn effective features, but rather to be optimized into a completely collapsed model. As a result, the model collapses completely, which outputs average values $\mathbb{E}_{x_m} x_m$ for any inputs, as shown in Fig .1(a).

To prevent model complete collapse in MIM, a novel masking strategy, the Hybrid Mask Pattern (HMP) masking strategy is designed for multi-modal MRI to directly improve $Var(x_m)$. As shown in Fig. 3, the HMP includes three types of masks: 1) **Modal Mask**, which randomly masks one modal. This encouraged the encoder to learn the information of the missing modal during reconstruction. 2) **Position Mask**, where positions are randomly selected, and the same positions are used to mask all modalities. The content information between different modalities of multi-modal MRI is highly coupled [30]. Image patches from different modalities at the same positions can be regarded as independent from each other. Therefore, the image patches can be masked or reserved independently. 3) **Patch Mask**, which complementarily selects patches of one position on different modalities. This forces the model to learn the semantic connection of each modal by reconstructing masked patches.

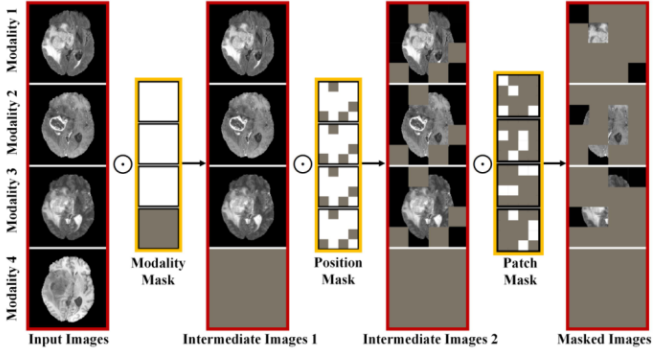

**Fig. 3.** Overview of the proposed masking strategy HMP. The input images are masked by HMP in three consecutive steps: 1) Modal Mask, 2) Position Mask, and 3) Patch Mask.

In this way, the content of the original image is maximally maintained while increasing the variance of the masked image, i.e., $Var(x_m)$ above the normally converged loss value, as shown in Fig. 2. This ensures that when the model falls into complete collapse during training, the loss has not yet converged. As a result, the parameters of the model can still be optimized along the gradient, allowing it to learn semantic features for reconstructing masked images and avoiding complete collapse.

### B. Avoiding Dimensional Collapse by Pyramid Barlow Twins Module

In the preceding section, we discussed the issue of total collapse. However, the model remains susceptible to dimensional collapse. This section examines the factors leading to MIM occurrence by considering the joint probability distribution of image patches and presents a solution.

MIM builds a connection between two complementary views $x_u$, $x_m$ through the reconstruction task. It can be modeled by a graph, as shown in Fig. 4 (a). The nodes of the graph are $x_u$ or $x_m$. The edge weight $\omega_{x_u x_m}$ of the graph between a pair of $x_u$, $x_m$ is defined as their joint probability $\mathcal{M}(x_u, x_m) = \mathbb{E}_x \mathcal{M}(x_u, x_m | x) = \mathbb{E}_x \mathcal{M}(u^1, u^2, \ldots, u^{n*(1-\rho)}, m^1, m^2, \ldots, m^{n*\rho} | x)$. We find that not only is there an edge when $x_m$ is the fully reconstructed target of $x_u$, but there is also an edge if two volumes have shared masked and unmasked patches. Consider two volume $x^i$, $x^j$, if $x_u^i \cap x_u^j \neq \phi$, $x_m^i \cap x_m^j \neq \phi$, then $\omega_{x_u^i x_u^j} = \mathcal{M}(x_m^i, x_m^j) \neq 0$, there is an edge between $x_m^i$ and $x_m^j$. For a closer look, by reconstructing different inputs $x^i$, $x^j$ with same groups of patches $\{u^1, u^2, \ldots\}$, $\{m^1, m^2, \ldots\}$, MIM implicitly aligns the latent features $z^i = f(x_u^i)$ and $z^j = f(x_u^j)$ with jointly probability distribution together, as shown in Fig. 4 (b). Just like contrastive learning, MIM aligns two similar inputs in the latent space through the encoder. In conclusion, the MIM algorithm utilizes random masking to implicitly align groups of similar image patches, facilitating the learning of feature representations.

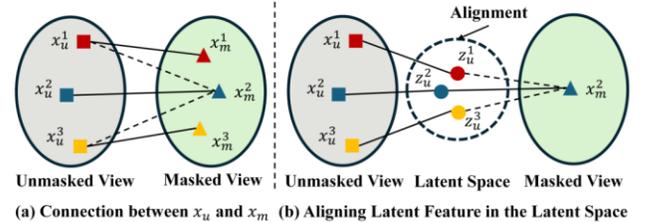

**Fig. 4.** (a) The graph is built by the connection between two views. Not only is there an edge between complementary views, but there is an edge if two volumes share a few masked and unmasked patches. (b) By this connection, latent features can be aligned in the latent space.

However, due to the absence of a negative pair definition, the model cannot ensure feature uniformity, causing dimensional collapse. This limits representation power, as the learned features lie in a low dimensional subspace.

To alleviate model dimensional collapse in MIM, the Pyramid Barlow Twins (PBT) module is proposed to explicitly add feature uniformity. As shown in Fig. 5, the PBT module comprises $L$ Barlow Twins [20] operating across different receptive fields. For the $l$-th Barlow Twins, the cross-correlation matrix $C^l$ between the latent features is computed. The latent features $z^l$ and $z_m^l$ are obtained by processing unmasked and masked multi-modal MR images through $\frac{12}{L} \times l$ -th Transformer [4] and reshape operation, respectively. It can be written as:

$$C_{ij}^l = \frac{\sum z^{l,i} z_m^{l,j}}{\sqrt{\sum \left(z^{l,i}\right)^2} \sqrt{\sum \left(z_m^{l,j}\right)^2}} \quad (4)$$

$$z^l = \text{Reshape}(\text{Transformer}^{\frac{12}{L} \times l}(x)), \ z^l = \{z^{l,0}, z^{l,1}, \cdots, z^{l,n}\} \quad (5)$$

$$z_m^l = \text{Reshape}(\text{Transformer}^{\frac{12}{L} \times l}(x_m)), \ z_m^l = \{z_m^{l,1}, z_m^{l,2}, \cdots, z_m^{l,n}\} \quad (6)$$

The loss function $\mathcal{L}_{PBT}$ follows:

$$\mathcal{L}_{PBT}^{total} = \sum_l \mathcal{L}_{PBT}^l \quad (7)$$

$$\mathcal{L}_{PBT}^l = \left(I^l - C^l\right)^2 = \sum_i \left(1 - C_{ii}^l\right)^2 + \sum_i \sum_{j \neq i} C_{ij}^{l\,2} \quad (8)$$

where $I^l$ is an identity matrix with the same size as $C^l$.

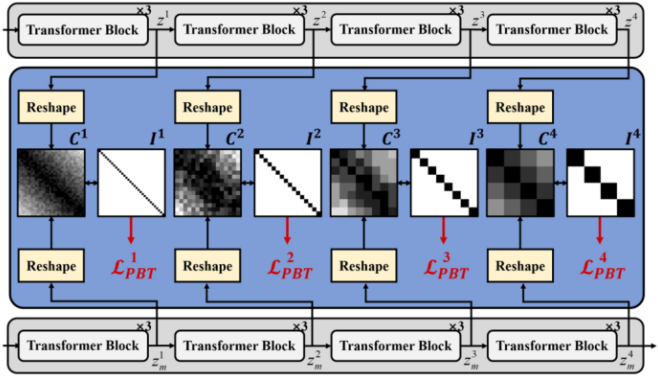

**Fig. 5.** The detail of the Pyramid Barlow Twins (PBT) module. The PBT module aims to align the semantic representations of input and masked images by implementing feature pyramid upon the encoder.

Intuitively, by trying to equate the diagonal elements of the cross-correlation matrix to 1, makes the embeddings of same position patches close. By trying to equate the non-diagonal elements of the cross-correlation matrix to 0, the embedding uniformity is explicitly added by reducing the redundancy between different position patches.

*C. Enhanced MIM*

Within this section, we introduce Enhanced MIM (E-MIM), a method that effectively merges the HMP masking strategy with the PBT module to alleviate instances of model collapse, encompassing both complete and dimensional collapse, that may arise during pretraining on multi-modal MRI datasets.

In E-MIM (as shown in Fig. 6), the input image $x$ is masked by HMP to get $x_m$, firstly. $x$ and $x_m$ are then separately entered into vision Transformer-based encoders that share weights. Note that the encoders are responsible for modeling latent feature representations of the masked patches, which are used to reconstruct the original image patches in the masked area. As a standard ViT, the encoders of E-MIM embed patches through linear projection with the incorporation of positional embeddings and process the resulting features via a series of Transformer blocks. The shared weights of the encoders operate on both the masked and unmasked patches of the entire dataset, producing embeddings and for the input and masked images, respectively. Thirdly, the embedding of each transformer block is fed to a PBT module for explicitly adding feature uniformity at various scales. Lastly, $z_m^L$ are reconstructed by a decoder. The input to the decoder is the full set of tokens consisting of encoded visible patches. A lightweight decoder reduces computational complexity and increases the ability of the encoder to learn more generalizable representations that the decoder can quickly grasp, translate, and convey [29]. The encoder is more critical because the decoder is only used during pre-training to perform the image reconstruction task (only the encoder is used to produce image representations for recognition). Therefore, the entire decoder is replaced with a single output projection $g(\cdot)$ in E-MIM.

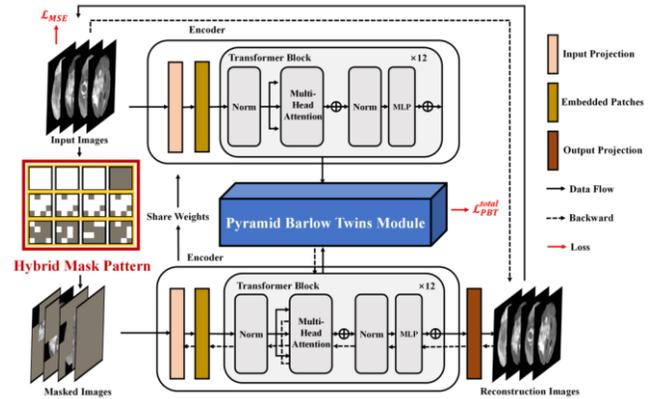

**Fig. 6.** Overview of the proposed Enhanced MIM (E-MIM). The input images and masked images masked by HMP are fed to the share weights encoder respectively to reconstruct masked image for pretraining the encoder.

To pretrain the encoder for multi-modal MRI datasets by reconstructing masked images, given the multi-modal MR image $x$, the Transformer encoder $Transformer(\cdot)$ and the output projection $g(\cdot)$, the MSE loss can be written as:

$$\mathcal{L}_{MIM} = \left| x - g\left(Transformer(x_m)\right) \right|^2 \quad (9)$$

In summary, the E-MIM is based on a MIM consisting of a PBT module Therefore, the overall E-MIM is optimized by a joint loss function:

$$\mathcal{L}_{overall} = \mathcal{L}_{MIM} + \mathcal{L}_{PBT}^{total} \quad (10)$$

IV. EXPERIMENTS

*A. Dataset*

**BraTS2023:** The multi-modal MR images of glioma patients are derived from the multi-modal brain tumor segmentation (BraTS2023) challenge [1]. Genomic information was available from The Cancer Genome Atlas and IDH mutation status, FLAIR, T1, T1ce, and T2 modalities were assessed. For the glioma segmentation task, the final training dataset included 5004 preoperative MR images from 1251 subjects; the test dataset included 219 unlabeled subjects. Glioma segmentation was submitted to BraTS2023 to determine the segmentation accuracy for various algorithms. In the training dataset, only 369 subjects have IDH mutation, including 76 subjects with low-grade glioma (LGG) and 293 subjects with high-grade glioma (HGG). Thus, 299 subjects with IDH mutation status were used as training data, and 70 subjects with IDH mutation status are used as test data for IDH classification tasks.

**PI-CAI:** The full dataset used for the PI-CAI [43] public dataset comprises a cohort of 1500 prostate MRI exams, curated from three Dutch and one Norwegian center. Imaging consisted of the following sequences: axial T2-weighted imaging (T2W), diffusion-weighted imaging (DWI), and apparent diffusion coefficient maps (ADC). Out of the 1500 cases shared in the dataset, 1075 cases have International Society of Urological Pathology (ISUP) ≤1, and 425 cases have ISUP ≥2. The official website also provides the segmentation labels on the transition zone (TZ) and peripheral zone (PZ). In the experiments, 1200 were used as training data and 300 subjects as test data by random data splitting.

**LungasMRI:** The LungasMRI acquisition is performed using the method previously reported by our group [34]. A total of 85 subjects were enrolled including 43 healthy volunteers and 42 patients with lung disease such as chronic obstructive pulmonary disease (COPD), asthma and so on. All subjects provided informed consents, and the experiments were approved by the local Institutional Review Board (IRB). The parameters for $^1$H imaging are as follows: repetition time/echo time (TR/TE) = 2.4/ 0.7 ms, matrix size = 96 × 96, number of slices = 24. The parameters for $^{129}$Xe imaging are as follows: TR/TE = 4.2/1.9 ms, matrix size = 96×96, number of slices = 24. Professional doctors segmented the $^1$H and $^{129}$Xe images to obtain thoracic cavity mask and ventilation mask using $^1$H and $^{129}$Xe MR images. In the experiments, 68 subjects were used as training data and 17 subjects as test data by randomly data splitting.

### B. Implementation Details

All the models are implemented in PyTorch. MONAI is used for data transformation and loading. ViT-based architecture [16] is used as the standard encoder backbone. For the supervised baseline of the segmentation task, we employ a batch size of 4, the Adam optimizer, and a learning rate of 0.0003 with a weight decay of 0.05 based on a linear warmup-up to 300 epochs and a cosine annealing scheduler. Training is conducted on four NVIDIA 3090 GPU for a total of 1000 epochs.

Due to the differences between segmentation and classification tasks, different decoders are used for each task. For the segmentation tasks, UNETR [4] is adopted as the default supervised baseline in the study, which is one of the SOTA models in medical imaging segmentation. Specifically,

UNETR is a U-shaped architecture employing a ViT as the encoder backbone and a convolutional upsampling decoder following the U-Net [2] design. For the classification tasks, self-supervised learning is widely used to evaluate the quality of pretraining by linear probing: the parameters of the encoder are fixed, and then a linear classifier is added to classify the images. Therefore, after adding the main encoder based on ViT, the projection head of one classification task is added as the classification header, the weights of the encoder are frozen and its pretraining, and the parameters of projection head are only updated under the fine-tune for the classification task.

### C. Comparison with the SOTA Methods

Extensive experiments were conducted to compare the E-MIM with state-of-the-art (SOTA) methods [3]-[8],[12],[18]-[20],[32],[44]-[46] on the three multi-modal datasets, as shown in Table I and Table II. To enable fair comparison, all the pretraining models adopted a Vision Transformer encoder with comparable parameters and were trained with the 3D images as input for pretraining, and the trained models were transferred to the downstream segmentation and classification tasks.

*1) Reconstruction Task on Pretraining Stage*

MAE [19] and E-MIM were used to reconstruct masked multi-modal MRI in pretraining stage. Fig. 7 (a) shows that MAE yielded reconstructed images with "averaging" effect of voxel signal intensities. However, our approach is capable of recovering high level memantic information such as the sharp contour of the brain and glioma. This suggests that E-MIM mitigates model complete collapse in the pretraining stage and thus facilitates thedownstream segmentation tasks. The singluar values and effective rank indicate that E-MIM can prevent model dimensional collapse, as shown in Fig. 7 (b) and Fig. 7 (c).

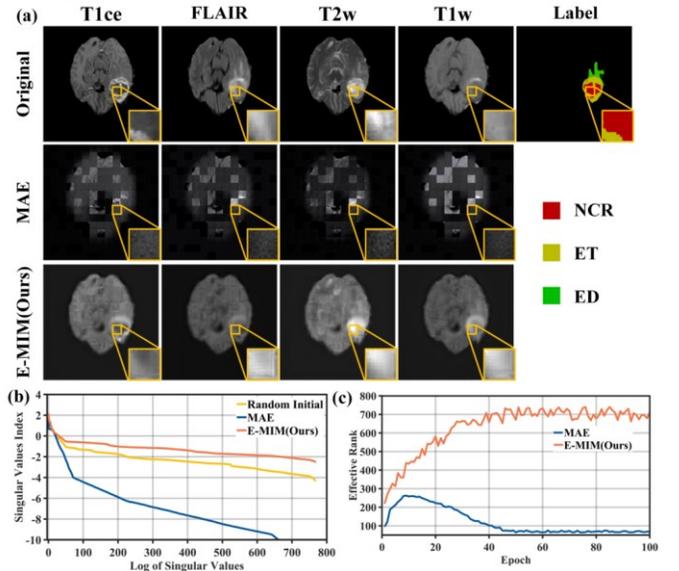

Fig. 7. (a) Visual comparison of reconstructed results produced by different MIM methods for the BraTS2023 dataset. As the original images are all 3D volumes, reconstructed images are shown in the form of slices. (b) Comparison of the singular values of learned

features with randomly initialized parameters, MAE, and E-MIM. (c) The changing process of the effective rank of the encoded features trained with MAE and E-MIM.

TABLE I
SEGMENTATION RESULTS OF THE PROPOSED METHOD AND OTHER EXISTING SOTA METHODS ON THREE MULTI-MODAL DATASETS

| Schemes | | BraTS2023 Dataset | | | | | | PI-CAI Dataset | | LungasMRI Dataset | |
|---|---|---|---|---|---|---|---|---|---|---|---|
| Method | Pretrain (w/o) | Dice (mean ± std) (%) | | | HD95 (mean ± std) (voxel) | | | Dice (mean ± std) (%) | | Dice (mean ± std) (%) | |
| | | WT | TC | ET | WT | TC | ET | TZ | PZ | TC | VD |
| 3D UNet [3] | — | 72.68±17.58 | 57.54±39.43 | 53.20±36.64 | 22.36±21.02 | 41.97±59.51 | 59.16±93.67 | 79.86±9.40 | 74.30±21.85 | 88.50±8.33 | 81.85±12.66 |
| DMF Net [4] | — | 79.18±15.74 | 61.70±37.18 | 58.67±31.85 | 21.53±25.97 | 37.38±52.11 | 48.37±87.96 | 83.25±9.81 | 73.94±18.04 | 90.77±9.76 | 82.08±11.80 |
| nnUNet [5] | — | 81.29±15.61 | 64.77±35.75 | 58.90±32.14 | 21.77±24.39 | 35.26±55.83 | 52.33±98.50 | 85.59±6.42 | 78.68±12.35 | 91.20±8.67 | 82.98±12.18 |
| TransUNet [6] | — | 85.84±13.93 | **81.61±22.57** | 76.63±26.57 | 12.35±28.32 | **8.91±17.38** | 19.68±55.50 | 87.72±7.30 | **81.72±6.42** | 93.25±4.88 | **87.92±8.52** |
| TransBTS [7] | — | 86.49±11.31 | 79.10±18.09 | **78.68±21.45** | 11.20±14.97 | 9.47±13.43 | **12.91±42.00** | 88.02±6.93 | 80.39±6.07 | 93.65±3.98 | 86.22±8.15 |
| UNETR [8] | — | **87.79±8.96** | 70.46±32.22 | 62.81±28.95 | **9.46±11.71** | 23.74±31.29 | 36.23±74.10 | **88.24±6.00** | 79.95±10.70 | **94.08±3.55** | 82.44±11.54 |
| Moco [12] | ✓ | 88.02±9.87 | 71.12±30.33 | 62.60±28.72 | 8.77±14.44 | 20.49±21.23 | 42.61±71.07 | 89.12±6.19 | 80.61±6.90 | 95.12±3.31 | 83.01±10.22 |
| Barlow Twins [20] | ✓ | 88.82±7.74 | 72.41±32.03 | 65.85±24.15 | 8.94±13.65 | 20.81±18.44 | 34.45±62.16 | 89.96±5.30 | 79.82±8.82 | 96.78±3.23 | 83.82±9.05 |
| MAE [19] | ✓ | 89.16±5.85 | 77.72±16.01 | 67.75±21.30 | 7.97±13.14 | 11.91±29.91 | 29.68±89.24 | 90.54±3.67 | 80.04±6.14 | 96.15±3.30 | 86.71±8.32 |
| SimMIM [18] | ✓ | **90.64±4.98** | 75.22±28.85 | 75.75±25.94 | **7.21±11.21** | 15.47±44.46 | 18.91±65.70 | 91.11±3.67 | 81.85±7.18 | 97.21±2.33 | 86.54±8.00 |
| E-MIM(Ours) | ✓ | 89.31±5.60 | **85.00±16.74** | **81.01±19.32** | 8.25±12.37 | **7.02±10.44** | **8.00±36.15** | **91.19±3.69** | **82.59±6.13** | **97.25±2.82** | **91.23±5.94** |

*2) Segmentation Task*

The segmentation results are shown in Table I. To quantitatively evaluate the performance of the models on segmentation tasks, the Dice similarity score (Dice) and the 95% Hausdorff distance (HD95) are employed. The segmentation performance of the methods without pretraining is in the top 6 rows. The segmentation performance of pretraining model is in the bottom 5 rows. Experimental results show that a model outperforms the comparative methods in the segmentation tasks. Particularly, the proposed E-MIM achieves a mean Dice of 89.71%, 85.02%, 81.29% for the whole tumor (WT), tumor core (TC), enhancing tumor (ET) on BraTS2023. The mean Dice for the transition zone (TZ) and peripheral zone (PZ) on PI-CAI are 91.19% and 82.59%, respectively, and the Dice for the thoracic cavity (TC) and ventilation defect (VD) region on LungasMRI is 97.25% and 91.23%, respectively. Most of the measurements obtained by the proposed method are the top among all the self-supervised segmentation methods. For the glioma segmentation task, the improvement in segmentation accuracy of TC and ET is most significant, with an increase of 14.56% and 18.48% in Dice compared to UNETR without pretraining. Furthermore, the qualitative results of E-MIM and other comparison approaches in three datasets are shown in Fig. 8, Fig. 9, and Fig. 10 respectively. The proposed E-MIM different methods for the BraTS2023 dataset. The glioma edema, tumor core and gangrene are presented in green, yellow, and red, respectively.

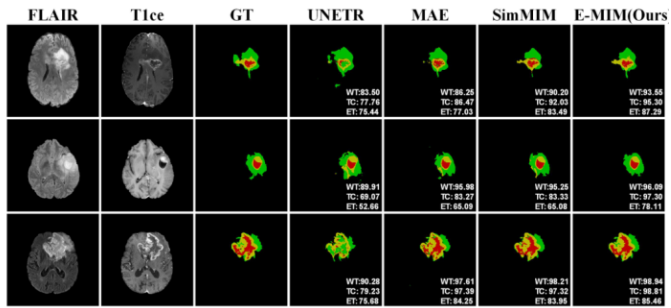

**Fig. 8.** Visual comparison of segmentation results produced by different methods for the BraTS2023 dataset. The glioma edema, tumor core and gangrene are presented in green, yellow, and red, respectively.

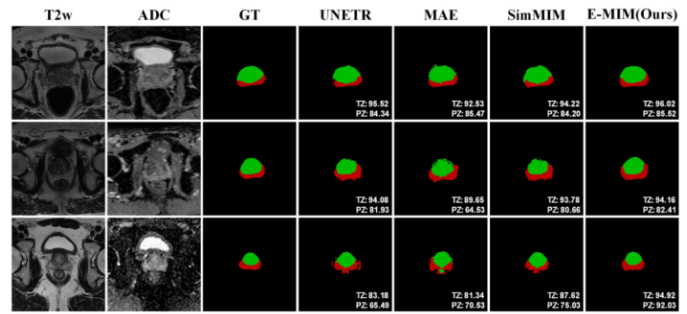

**Fig. 9.** Visual comparison of segmentation results produced by different methods for the PI-CAI dataset. The prostate transitional zone and zone of the prostate are presented in green and red, respectively.

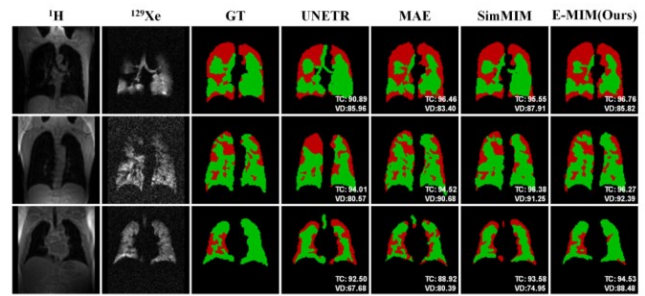

**Fig. 10.** Visual comparison of segmentation results produced by different methods for the LungasMRI dataset. The ventilation region and ventilation defect region are presented in green and red, respectively.

can produce more accurate segmentation results in most cases. Particularly, the edges of ET in Fig. 8 and TC in Fig. 10 are not misclassified, and the shape of PZ in Fig. 9 is complete.

*3) Classification Task*

The classification results are shown in Table II. For the classification tasks, area under the curve (AUC) and accuracy (Acc) are used for the quantitative evaluation. The E-MIM achieves remarkable classification performance for the classification task. Specifically, the E-MIM achieved an AUC of 72.46% for IDH genotyping, 84.60% for prostate cancer grading, and 76.88% for lung disease classification. Compared with the SOTA pretraining methods, the proposed method achieves the best classification AUC. In addition, the model can approximate the supervised training model by training the classification head only, which reflects the good feature extraction ability.

TABLE II
CLASSIFICATION RESULTS OF THE PROPOSED METHOD AND OTHER EXISTING SOTA METHODS ON THREE MULTI-MODAL DATASETS

| Schemes | | BraTS2023 Dataset | | PI-CAI Dataset | | LungasMRI Dataset | |
|---|---|---|---|---|---|---|---|
| Method | Pretrain (w/o) | AUC (%) | Acc (%) | AUC (%) | Acc (%) | AUC (%) | Acc (%) |
| ResNet50[44] | — | 76.42 | 73.91 | 86.52 | 83.67 | 78.69 | 70.59 |
| SeNet101[45] | — | 79.21 | 76.81 | 87.83 | 85.00 | 82.67 | 82.35 |
| DenseNet121[32] | — | 79.40 | 76.81 | 87.49 | 85.67 | 82.13 | 82.35 |
| ViT [46] | — | **80.04** | **79.71** | **91.28** | **90.33** | **88.91** | **88.24** |
| Moco | ✓ | 64.06 | 63.77 | 76.57 | 73.00 | 62.50 | 58.82 |
| Barlow Twins | ✓ | 68.38 | 65.22 | 78.28 | 76.33 | 60.63 | 58.82 |
| MAE | ✓ | 71.70 | 69.56 | 82.33 | 81.00 | 67.02 | 64.71 |
| SimMIM | ✓ | 70.50 | 68.12 | 83.52 | 81.67 | 68.38 | 64.71 |
| E-MIM(Ours) | ✓ | **72.46** | **71.01** | **84.60** | **82.00** | **78.88** | **70.59** |

*4) Semi-supervised Segmentation Task*

Based on dataset splitting, 20%, 40%, 60%, 80% and 100% of the labeled data are taken from the training data. Semi-supervised segmentation task is designed by using unlabeled data for E-MIM pre-training and using labeled data for supervised segmentation training (shown in Fig. 11). The dashed line represents the result of experiments using all labeled data without E-MIM pre-training weights. This exemplifies the potential impact of our approach in resource-constrained environments and its potential utility in field of multi-modality MRI analysis.

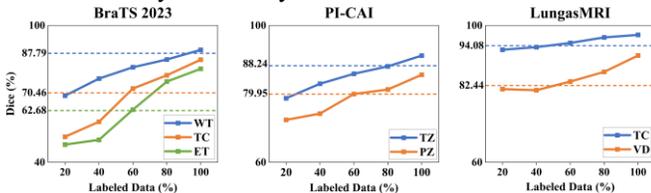

**Fig. 11**. An illustration of how E-MIM advances the semi-supervised learning on segmentation task. 20%, 40%, 60%, 80%, and 100% labeled data are used to train the segmentation model with E-MIM pre-training model.

*D. Ablation Study*

*1) Evaluation of Masking Strategy*

To demonstrate the effectiveness of the masking strategy, relevant ablation experiments are conducted. As shown in Table III, using the position mask and patch mask to reconstruct images during pretraining can effectively enhance the segmentation accuracy of TC and ET in BraTS2023, PZ in PI-CAI, and VD in LungasMRI. It has been indicated that, modal mask can improve the segmentation and classification accuracy of targets with different sizes, while patch mask can substantially improve the performance of small lesions.

TABLE III
EVALUATION OF PROPOSED MASKING STRATEGY ON THREE DATASETS

| Mask Strategy | | | BraTS2023 Dataset | | | | PI-CAI Dataset | | | LungasMRI Dataset | | |
|---|---|---|---|---|---|---|---|---|---|---|---|---|
| | | | Segment | | | Classify | Segment | | Classify | Segment | | Classify |
| Modal Mask | Position Mask | Patch Mask | Dice (mean) (%) | | | AUC (%) | Dice (mean) (%) | | AUC (%) | Dice (mean) (%) | | AUC (%) |
| | | | WT | TC | ET | | TZ | PZ | | TC | VD | |
| ✓ | | | 87.52 | 71.08 | 63.21 | 64.51 | 89.20 | 80.73 | 82.72 | 94.78 | 82.48 | 67.01 |
| ✓ | | | 89.16 | 77.72 | 67.75 | 71.70 | 90.54 | 80.04 | 82.33 | 96.15 | 86.71 | 67.02 |
| | ✓ | | 88.61 | 80.54 | 70,56 | 70.66 | 90.14 | 82.92 | 83.21 | 96.09 | 88.67 | 67.06 |
| | | ✓ | 88.66 | 78.81 | 73.90 | 70.75 | 90.58 | 83.20 | 83.79 | 96.02 | 87.10 | 66.97 |
| ✓ | ✓ | | **89.55** | 80.17 | 77.29 | 71.77 | 90.32 | 82.99 | 84.27 | 97.03 | 88.72 | 67.34 |
| ✓ | | ✓ | 88.98 | 81.65 | 76.90 | 71.32 | 91.19 | 84.42 | 84.25 | 96.78 | 90.11 | 67.12 |
| | ✓ | ✓ | 89.06 | 83.41 | 80.19 | 70.06 | 90.75 | 84.44 | 84.27 | 96.54 | 89.91 | 67.00 |
| ✓ | ✓ | ✓ | 89.31 | **85.00** | **81.01** | **72.46** | **91.19** | **85.59** | **84.60** | **97.25** | **91.23** | **78.88** |

*2) Evaluation of Pyramid Barlow Twins Module in Different Level*

The proposed PBT module facilitates the model to learn the relationship between modalities and forces the encoders to learn reconstruction features at different vision scale. To analyze the effect of PBT at different levels, the scheme is adopted at each level in the pretrained encoders. As shown in Table IV, the performance of the downstream tasks improves with levels and reaches the maximum at level=4. Considering the accuracy and quantity of training parameters, level = 4 is chosen as the optimal parameter of the number of transformer layers.

## V. DISCUSSION

In this study, we separately explore and analyze the causes of model complete collapse and model dimensional collapse in self-supervised training for multi-modal MRI. Based on this, we propose E-MIM for multi-modal MRI self-supervised learning. We design a masking strategy, HMP, to increase the lower bound of the training loss, providing a way to avoid model complete collapse and enhance the pretraining effect. Additionally, we develop a PBT module to explicitly improve feature uniformity at different visual scales to prevent dimensional collapse. Experiments show that E-MIM can effectively avoid model collapse, enhance pretraining performance, and improve the accuracy of downstream segmentation and classification tasks.

TABLE IV
EVALUATION OF E-MIM USING PYRAMID BARLOW TWINS MODULE IN VARIOUS LEVEL

| Level | BraTS2023 Dataset | | | | PI-CAI Dataset | | | LungasMRI Dataset | | |
|---|---|---|---|---|---|---|---|---|---|---|
| | Segment | | | Classify | Segment | | Classify | Segment | | Classify |
| | Dice (mean) (%) | | | AUC (%) | Dice (mean) (%) | | AUC (%) | Dice (mean) (%) | | AUC (%) |
| | WT | WC | ET | | TZ | PZ | | TC | VD | |
| 1 (input) | 88.81 | 84.14 | 80.76 | 71.27 | 90.87 | 85.22 | 84.09 | 96.83 | 90.99 | 74.49 |
| 2 ($6^{th}$, $12^{th}$-layer) | 88.85 | 84.51 | 80.81 | 71.71 | 90.81 | 85.28 | 84.02 | 96.01 | 91.12 | 74.70 |
| 3 ($4^{th}$, $8^{th}$, $12^{th}$-layer) | 89.24 | 84.77 | 80.89 | 72.30 | 91.09 | 85.54 | **84.78** | 97.15 | 90.91 | 77.98 |
| 4($3^{rd}$, $6^{th}$, $9^{th}$, $12^{th}$-layer) | 89.31 | **85.00** | **81.01** | **72.46** | **91.19** | **85.59** | 84.60 | **97.25** | **91.23** | **78.88** |
| 12(all layer) | **89.70** | 83.15 | 79.50 | 69.54 | 90.24 | 83.45 | 83.71 | 96.66 | 90.60 | 67.11 |

The HMP directly improves the variance of masked images to reduce the risk of model complete collapse. Current self-supervised medical image analysis methods mostly treat multi-modal MRI as an indivisible whole and ignore the relationship between paired modalities. Although images from different modalities for the same sample present their exclusive features, they inherit some global content structures. For instance, the underlying anatomical structure of the brain is shared by all modalities in the BraTS2023 dataset. At present, the modal translation task has been proven to be beneficial for segmentation tasks. This means that reconstructing missing full modal or missing modal patches as the pretext task is significant for downstream tasks. For example, small lesion areas are easy to ignore by Vit-based encoders, while effective improvement is to use a modal mask and patch mask (as shown in Table III by mask reconstruction. In addition, representations of multi-modal MRI through HMP masking can be better fused than other pretext tasks, like [16]-[18]. This is because learning the semantic connections among modalities can directly reconstruct the masked modal and image patches [47]. HMP forces the model to explicitly capture correlations across different modalities, improving the robustness of the model even with a small training dataset.

To prevent model dimensional collapse, we introduce a PBT module in E-MIM to regularize the model by explicitly adding feature uniformity at different vision scales in the latent space. In the PBT module, the performance of the downstream segmentation tasks is substantially enhanced with level=4. This may be attributed to the high efficiency of the parameter, reflecting the number of jump connection layers in UNETR. However, when the PBT was added to all levels in the encoder, there was a noticeable decrease in the performance of the downstream classification and segmentation tasks, which may be attributed to the large number of training parameters that causes difficulties for the gradient to go backward.

We further evaluated the performance of E-MIM by reducing the amount of the labeled data. Fig. 11 demonstrates the comparison results of fine-tuning using a subset of three datasets. Using only 60% of the labeled dataset for supervised training, experiments with E-MIM pretraining weights achieved similar performance compared to training from scratch on small regions like TC, ET, PZ, and VD. Our findings suggest promising future implications and potential extensions of this work for other medical imaging tasks. Compared with traditional semi-supervised learning, consistency-based semi-supervised is built on a consistent data distribution, which may not suit datasets from multiple data centers. Pseudo label-based semi-supervised learning methods may produce error amplification and increase the risk of training instability. In contrast, the self-supervised learning is used to learn universal features through a pretext task, which can reduce errors from data bias.

We acknowledge several study limitations. For example, regarding the generality of the method to different tasks, the Modal Mask strategy in the HMP randomly masks a whole modal for the input multi-modal MRI. However, each modality varies in importance for downstream tasks; for example, T1ce is more important than T1 for glioma segmentation. The adoption of the proposed mask strategy for use in specific lesions, organs, and downstream tasks warrants further refinement.

## VI. CONCLUSION

In this study, we discover and address the issue of model collapse on multi-modal MRI. The main area of focus is on complete and dimensional collapse, with an aim to clarify their causes and explore solutions. We attribute the model complete collapse to the lower variance of masked patches in multi-modal MRI compared to the loss during normal convergence. To address this, a hybrid mask pattern is proposed to increase the variance and prevent model complete collapse. An interpretable model for MIM is developed based on the joint probability distribution of image patches sets. The dimensional collapse observed in MIM models results from the lack of feature uniformity. To tackle this issue, a pyramid barlow twins module is introduced to explicitly improve the uniformity of model features and prevent model dimensional collapse. Experimental results show that the proposed method effectively prevents model complete collapse and model dimensional collapse. It results in an enhanced performance in downstream segmentation and classification tasks. This approach provides a reliable strategy for leveraging unlabeled data in self-supervised learning to improve performance in downstream segmentation and classification tasks.